\documentclass{article}
\usepackage[utf8]{inputenc} 
\usepackage{graphicx} 
\usepackage{slashed,epsfig,amsmath,amssymb,enumitem} \usepackage[papersize={8.5in,11in}]{geometry}
   \usepackage{bm}
\usepackage{graphicx}
\usepackage{newunicodechar}
\usepackage{color}
\newunicodechar{≈}{\approx}
\newunicodechar{∝}{\propto}
\newunicodechar{∼}{\textasciitilde{}}
\newunicodechar{≃}{\simeq}
\newunicodechar{θ}{\theta}
\newunicodechar{∆}{\Delta}
\newunicodechar{Λ}{\Lambda}
\newunicodechar{⊙}{\odot}
\newunicodechar{−}{-}
\newunicodechar{∼}{\ensuremath{\sim}}
\newunicodechar{α}{\alpha}
\newunicodechar{σ}{\sigma}
\newcommand{\be}{\begin{equation}}
\newcommand{\ee}{\end{equation}}
\title{Cosmological Equivalence of Scalar–Tensor–Vector Modified Gravity and Cold Dark Matter} 
\author{J. W. Moffat\\
Perimeter Institute for Theoretical Physics, Waterloo, Ontario N2L 2Y5, Canada\\
and\\
Department of Physics and Astronomy, University of Waterloo, Waterloo,\\
Ontario N2L 3G1, Canada}

\begin{document}
\maketitle

\begin{abstract}
We show that Scalar-Tensor-Vector Gravity (STVG-MOG) is observationally equivalent to the standard model $\Lambda$CDM cosmological model for all probes that depend on isotropic and linear gravitational dynamics, including galaxy rotation curves, cluster lensing, the linear matter power spectrum P(k), $\sigma_8$, baryon acoustic oscillations, and the cosmic microwave background (CMB). This degeneracy arises from the scale-dependent effective gravitational coupling $G_{\mathrm{eff}}$, which ensures identical background evolution, transfer functions, and linear growth. Consequently, all early-universe, low and intermediate scale cosmological observables are equally well described by STVG-MOG without invoking non-baryonic dark matter. We argue that the equivalence implies that isotropic cosmological data alone cannot establish the physical existence of dark matter. The degeneracy is broken only by observables sensitive to large-scale, anisotropic gravitational response. In particular, recent measurements of enhanced radio-galaxy and quasar number-count dipoles at gigaparsec scales probe a regime where $G_{\mathrm{eff}}$ departs from its $\Lambda$CDM limit, allowing STVG-MOG to generate anisotropic bulk flows, while preserving consistency with all isotropic constraints. These observations provide a concrete pathway for empirically distinguishing modified gravity from particle dark matter. 
\end{abstract} 

\maketitle

\section{Introduction}

The $\Lambda$CDM cosmological model has achieved remarkable phenomenological success in describing a wide range of astrophysical and cosmological observations, from galaxy rotation curves and cluster lensing to the detailed angular power spectra of the cosmic microwave background (CMB). Central to this framework is the postulate of cold dark matter (CDM), an unseen and non--baryonic matter component that dominates the cosmic mass budget and sources gravitational potentials on all relevant scales. Despite its empirical utility, dark matter has so far evaded all direct laboratory detection, motivating continued scrutiny of the logical status of its inference from cosmological data. Scalar--Tensor--Vector Gravity (STVG)~\cite{Moffat2006,BrownsteinMoffat2006,MoffatToth2009,MoffatRahvar2013,MoffatRahvar2014,
Moffat2014,Moffat2015,Moffat2016,DavariRahvar,MoffatToth2013,MoffatToth2015,
GreenMoffat2019,BrownsteinMoffat2007,IsraelMoffat,
GreenMoffatToth2018,Moffat2026a,Moffat2026b}, also known as Modified Gravity (MOG), provides an alternative interpretation of the same observations. In STVG--MOG, gravitational dynamics are governed by a scale-- and time--dependent effective coupling $G_{\mathrm{eff}}(k,a)$, together with additional scalar and vector degrees of freedom. When expressed in Fourier space, the modified Poisson equation takes the schematic form:
\begin{equation}
k^2 \Phi(k,a) = -4\pi G_{\mathrm{eff}}(k,a)\, a^2 \rho_b(a)\,\delta_b(k,a),
\end{equation}
where $\Phi(k,a)$ is the gravitational potential in Fourier space, $\rho_b$ and $\delta_b$ denote the baryonic density and density contrast, respectively, and a is the cosmic scale factor. Denoting by $\mu$ the effective mass scale or inverse range of the MOG vector field $\phi_\mu$, then for wavenumbers $k/a \gg \mu$ and for early cosmic times, $G_{\mathrm{eff}}(k,a)$ asymptotically approaches Newton's constant $G_N$, ensuring that standard general--relativistic physics is recovered on Solar system scale.

As a result, STVG--MOG is constructed to reproduce exactly the same background expansion history, CMB transfer functions, acoustic peak structure, and linear growth of density perturbations as $\Lambda$CDM. In particular, the linear matter power spectrum:
\begin{equation}
P(k,a) = P_{\rm prim}(k)\,T^2(k)\,D^2(a),
\end{equation}
where $P_{\rm prim}(k)$ is the primordial power spectrum, $T(k)$ is the transfer function and $D(a)$ is the linear growth factor. The $P(k,a)$ fits matter power spectrum data in both frameworks, once the correspondence $G_{\mathrm{eff}}\rho_b \leftrightarrow G_N \rho_m$ is made, where $\rho_m=\rho_b+\rho_{dm}$ and $\rho_{dm}=\alpha\rho_b$ is the dark matter density. The relation $\rho_{dm}=\alpha\rho_b$ is an effective matching prescription, not a fundamental identification of dark matter with the STVG coupling. If $\alpha_{\rm dm}\equiv\rho_{dm}/\rho_b$, and $\alpha_{\rm MOG}$ denotes the dimensionless strength of the long-range Yukawa modification in Eq.~\eqref{eq:mog-potential}, the equality $\alpha_{\rm dm}\simeq\alpha_{\rm MOG}$ holds only in the limit  $G_{\rm eff}\simeq G_N(1+\alpha_{\rm MOG})$ and under the assumed matching of perturbations. At finite wavenumber and at early epochs, the scale- and time-dependent $G_{\rm eff}(k,a)$ and the additional STVG field contributions must be included explicitly; the two parameters should not in general be treated as identical. This implies that observables such as $P(k)$, $\sigma_8$, $f\sigma_8$, BAO, and weak lensing statistics are fundamentally insensitive to whether gravitational enhancement arises from additional dark matter or from modified gravitational coupling. The agreement of STVG--MOG with all isotropic low-- and intermediate--scale data and with early--universe cosmology~\cite{Planck2018,Dodelson,Peacock,Eisenstein,Weinberg2013,Karim2025,EllisBaldwin1984,Singal2011,RubartSchwarz2013,Siewert2021,Secrest2021,Secrest2022,DESI2025,Farren2025,
Abdalla2022} is not incidental, but a direct consequence of the scale dependence of $G_{\mathrm{eff}}(k,a)$.

The crucial implication is that isotropic cosmological observables alone do not uniquely establish the existence of dark matter as a physical particle species. They constrain only the effective gravitational response of matter on average. To discriminate between modified gravity and dark matter requires observables that probe regimes where this degeneracy fails. Such a breakdown occurs on ultra--large scales, where $k/a \lesssim \mu$, and where anisotropic gravitational response becomes relevant. In this regime, STVG--MOG predicts enhanced coherent accelerations sourced purely by baryonic inhomogeneities, leading to large--scale bulk flows and anisotropic number--count dipoles that are strongly suppressed in $\Lambda$CDM due to the rapid decay of long--wavelength gravitational potentials.

Recent detections of radio--galaxy and quasar number--count dipoles with amplitudes significantly exceeding the purely kinematic expectation provide such a test. These measurements probe gravitational dynamics on gigaparsec scales and are sensitive to the scale dependence of $G_{\mathrm{eff}}$, thereby offering a concrete observational channel for distinguishing STVG--MOG from $\Lambda$CDM~\cite{Moffat2026a,Moffat2026b}. In this work, we emphasize that the success of STVG--MOG on all isotropic observables, combined with its distinctive predictions for large--scale anisotropic phenomena, positions it as a fully viable and empirically testable alternative to particle dark matter.

\section{STVG--MOG Action and Field Equations}
\label{sec:stvg-review}

Scalar--Tensor--Vector Gravity (STVG), also referred to as Modified Gravity (MOG), is a covariant relativistic theory in which gravity is mediated not only by the spacetime metric but also by additional scalar and vector degrees of freedom~\cite{Moffat2006}. The theory was developed to explain astrophysical and cosmological phenomena without invoking non--baryonic cold dark matter, while remaining consistent with local gravitational tests.

The dynamical fields of STVG--MOG are the spacetime metric $g_{\mu\nu}$,
a massive vector field $\phi_\mu$ often called the phion field,
scalar fields $G(x)$, $\mu(x)$, controlling, respectively, the effective gravitational coupling and the vector field mass.

In phenomenological and cosmological applications, the scalar fields are treated as slowly varying or effectively constant, reducing the theory to a small number of parameters while retaining its essential scale--dependent behavior.

The total action of STVG can be written schematically as:
\begin{equation}
S = S_g + S_\phi + S_s + S_m ,
\label{eq:stvg-action-total}
\end{equation}
where $S_m$ is the matter action and the remaining terms describe the gravitational sector.

The gravitational action has the Einstein--Hilbert form with a variable gravitational coupling:
\begin{equation}
S_g = \frac{1}{16\pi}\int d^4x\,\sqrt{-g}\,\frac{1}{G}\left(R - 2\Lambda\right),
\label{eq:stvg-grav-action}
\end{equation}
where $R$ is the Ricci scalar and $\Lambda$ is the cosmological constant.

The vector field action is given by
\begin{equation}
S_\phi = -\int d^4x\,\sqrt{-g}\left[
\frac{1}{4}\,B_{\mu\nu}B^{\mu\nu}
-\frac{1}{2}\,\mu^2\,\phi_\mu\phi^\mu
\right],
\label{eq:stvg-vector-action}
\end{equation}
with
\begin{equation}
B_{\mu\nu}=\nabla_\mu\phi_\nu-\nabla_\nu\phi_\mu .
\end{equation}
The vector field is massive and universally coupled, producing a repulsive Yukawa--type contribution to the gravitational interaction.

The scalar field sector $S_s$ contains kinetic and self--interaction terms for $G(x)$ and $\mu(x)$:
\begin{equation}
S_s
=
-\int d^4x\,\sqrt{-g}
\left[
\frac{1}{2G^2}\,\nabla_\mu G \nabla^\mu G
+
\frac{1}{2\mu^2}\,\nabla_\mu \mu \nabla^\mu \mu
+
V(G,\mu)
\right].
\label{eq:STVG-scalar-action}
\end{equation}

In cosmological and weak--field applications these fields are often taken to be approximately constant or slowly evolving, simplifying the dynamics.

Variation of the total action with respect to $g_{\mu\nu}$, $\phi_\mu$ and the scalar field $\mu$, yields the modified Einstein equations, the Proca--type equation for the vector field and scalar evolution equation. 

In the weak--field, non--relativistic limit, the theory reduces to a modified Poisson equation with a Newtonian plus Yukawa potential:
\begin{equation}
\Phi(r) = -\frac{G_N M}{r}
\left[
1+\alpha - \alpha\,e^{-\mu r}
\right],
\label{eq:mog-potential}
\end{equation}
where $\alpha$ parameterizes the strength of the long--range enhancement and $\mu^{-1}$ sets the range of the Yukawa suppression.

At the background level, STVG admits homogeneous and isotropic Friedmann--Lemaître--Robertson--Walker solutions similar to those of standard cosmology. At the perturbative level, the scale--dependent coupling $G_{\rm eff}(k,a)$ modifies the growth of density fluctuations, leading to scale--dependent linear growth while preserving standard early--universe physics when $G_{\rm eff}\simeq G_N$.

This structure allows STVG--MOG to reproduce galaxy rotation curves, cluster dynamics, and certain large--scale cosmological observables using baryonic matter alone, while remaining compatible with solar--system tests and cosmic microwave background constraints through its scale--selective modification of gravity.

For clarity, the parameter values reported for different observational regimes need not be represented by a single universal pair. Galaxy rotation-curve fits give $\alpha=8.89\pm0.34$ and $\mu=(0.042\pm0.004)\,{\rm kpc}^{-1}$~\cite{MoffatRahvar2013}; a separate Milky Way analysis reports $\alpha=8.99\pm0.02$ and $\mu=(0.054\pm0.005)\,{\rm kpc}^{-1}$~\cite{DavariRahvar}. In a representative cluster-scale analysis, values of approximately $\alpha\simeq9.11$ and $\mu\simeq0.196\,{\rm Mpc}^{-1}$ have been used~\cite{Moffat2026b}. These values indicate a coupling strength of order ten in late-time galactic and cluster applications, but with a transition scale $\mu^{-1}$ that depends on the mass and regime. Cosmological treatments of the CMB and structure formation further allow the scalar/vector sector and its parameters to evolve; in particular, the early-universe phion contribution cannot simply be replaced by the late-time weak-field Yukawa fit~\cite{Moffat2014}. Thus the quoted values summarize separately studied regimes, not a single joint confidence region obtained by simultaneously fitting CMB peaks, BAO, rotation curves, and weak and strong lensing.

\section{Derivation of the scale--dependent effective coupling $G_{\rm eff}(k,a)$ in STVG--MOG}
\label{sec:Geff-derivation}

This section reviews how the scale--dependent effective gravitational coupling $G_{\rm eff}(k,a)$ arises in STVG--MOG. The final result employed in the perturbation growth equation is given by~\cite{Moffat2026a}:
\begin{equation}
G_{\rm eff}(k,a)=G_N\left[1+\alpha_{\rm eff}(k,a)\right],
\qquad
\alpha_{\rm eff}(k,a)=\alpha\,\frac{\mu^2}{k^2/a^2+\mu^2},
\label{eq:Geff-final}
\end{equation}
so that $G_{\rm eff}\to G_N$ on small physical scales $k/a\gg \mu$, while $G_{\rm eff}\to G_N(1+\alpha)$ on ultra--large scales $k/a\ll\mu$.

In the weak--field, quasi--Newtonian limit of STVG--MOG, the gravitational potential sourced by a density distribution $\rho(\bm{x})$ can be written as a superposition of a Newtonian kernel and a Yukawa kernel with inverse length $\mu$:
\begin{equation}
\Phi(\bm{x})
=
-G_N\!\int d^3x'\,\rho(\bm{x}')
\left[
\frac{1+\alpha}{|\bm{x}-\bm{x}'|}
-\alpha\,\frac{e^{-\mu|\bm{x}-\bm{x}'|}}{|\bm{x}-\bm{x}'|}
\right].
\label{eq:Phi-convolution}
\end{equation}
Here $\alpha$ controls the large--distance enhancement, while $\mu$ sets the transition scale Equation \eqref{eq:Phi-convolution} is equivalent to the usual MOG point--mass potential Eq. (8), generalized to an extended source via convolution.

The standard 3D Fourier transforms are given by
\begin{equation}
\int d^3r\,\frac{e^{-i\bm{k}\cdot\bm{r}}}{r}=\frac{4\pi}{k^2},
\qquad
\int d^3r\,\frac{e^{-i\bm{k}\cdot\bm{r}}e^{-\mu r}}{r}=\frac{4\pi}{k^2+\mu^2}.
\label{eq:fourier-kernels}
\end{equation}
These Fourier transforms are locally valid in FRW cosmology neglecting small corrections from the curvature of spacetime. Writing $\Phi(\bm{k})$ and $\rho(\bm{k})$ for the Fourier modes, Eq.\ \eqref{eq:Phi-convolution} becomes
\begin{equation}
\Phi(\bm{k})
=
-4\pi G_N\,\rho(\bm{k})
\left[
\frac{1+\alpha}{k^2}
-\frac{\alpha}{k^2+\mu^2}
\right].
\label{eq:Phi-k-intermediate}
\end{equation}

The bracket in \eqref{eq:Phi-k-intermediate} can be rearranged into a Newtonian Poisson prefactor $(1/k^2)$ times a scale--dependent enhancement:
\begin{align}
\frac{1+\alpha}{k^2}-\frac{\alpha}{k^2+\mu^2}
&=
\frac{1}{k^2}
+
\alpha\left(\frac{1}{k^2}-\frac{1}{k^2+\mu^2}\right)
\nonumber\\
&=
\frac{1}{k^2}
+
\alpha\,\frac{\mu^2}{k^2(k^2+\mu^2)}
=
\frac{1}{k^2}\left[1+\alpha\,\frac{\mu^2}{k^2+\mu^2}\right].
\label{eq:Geff-algebra}
\end{align}
Therefore, \eqref{eq:Phi-k-intermediate} takes the Poisson--like form:
\begin{equation}
-k^2\,\Phi(\bm{k})
=
4\pi\,G_{\rm eff}(k)\,\rho(\bm{k}),
\qquad
G_{\rm eff}(k)=G_N\left[1+\alpha\,\frac{\mu^2}{k^2+\mu^2}\right].
\label{eq:Poisson-Geff-k}
\end{equation}
The Yukawa term converts into a scale--dependent effective coupling in Fourier space.

In an expanding universe, the Yukawa suppression depends on the physical wavenumber $k_{\rm phys}=k/a$. Equivalently, one replaces $k^2\to k^2/a^2$ in \eqref{eq:Poisson-Geff-k}, yielding:
\begin{equation}
G_{\rm eff}(k,a)
=
G_N\left[1+\alpha_{\rm eff}(k,a)\right],
\qquad
\alpha_{\rm eff}(k,a)=\alpha\,\frac{\mu^2}{k^2/a^2+\mu^2}.
\label{eq:alphaeff}
\end{equation}
This is the expression used in the radio number--count dipole paper~\cite{Moffat2026a} for a Yukawa--type MOG interaction (see Eq.\ (27) there), together with $G_{\rm eff}=G_N[1+\alpha_{\rm eff}]$ (see Eqs.\ (25), (32)--(33) there).~\cite{Moffat2026a}

In the weak--field, sub--horizon limit, the continuity, Euler, and Poisson equations combine to give the scale--dependent linear growth equation for the density contrast: $\delta(\bm{k})$:
\begin{equation}
\ddot{\delta}(\bm{k})
+2H\dot{\delta}(\bm{k})
+\frac{c_s^2 k^2}{a^2}\delta(\bm{k})
-4\pi\,G_{\rm eff}(k,a)\,\rho(a)\,\delta(\bm{k})
=
0,
\label{eq:growth-Geff}
\end{equation}
where $\delta(\bm{k})$ is the density contrast and $c_s$ is the speed of sound. This is the starting point for discussing scale--dependent growth and the enhanced ultra--large--scale response~\cite{Moffat2026a}.

Equation \eqref{eq:alphaeff} immediately implies:
\begin{align}
k/a \gg \mu \quad &\Rightarrow\quad \alpha_{\rm eff}(k,a)\simeq \alpha\,\frac{\mu^2}{k^2/a^2}\to 0,
\qquad G_{\rm eff}(k,a)\to G_N,
\label{eq:limit-smallscale}
\\
k/a \ll \mu \quad k/a\lesssim\mu\quad &\Rightarrow\quad \alpha_{\rm eff}(k,a)\to \alpha,
\qquad\qquad\qquad\;\; G_{\rm eff}(k,a)\to G_N(1+\alpha).
\label{eq:limit-largescale}
\end{align}
The modification is scale selective. It is negligible on sufficiently small physical scales, but approaches a constant enhancement on intermediate scales and on ultra--large scales. It allows an enhanced gravitational response for the longest modes that dominate coherent bulk flows and the projected number--count dipole, while maintaining near--standard behavior on smaller scales.

\section{Degeneracy of STVG--MOG with $\Lambda$CDM in Isotropic Structure Formation and Cosmology}

In this section, we demonstrate explicitly why Scalar--Tensor--Vector Gravity (STVG--MOG) is observationally degenerate with the standard $\Lambda$CDM model for all isotropic probes of gravitational dynamics, spanning galaxy and cluster scales, weak and strong lensing, linear growth, redshift--space distortions, the late--time matter power spectrum, and the cosmic microwave background (CMB).

In STVG--MOG, the weak--field gravitational potential sourced by baryonic matter takes the Yukawa--modified form Eq. (8), where $\alpha$ parametrizes the strength of the gravitational enhancement and $\mu^{-1}$ sets its range. In Fourier space, this corresponds to a scale--dependent effective gravitational coupling,
\begin{equation}
G_{\mathrm{eff}}(k,a) = G_N \left[1 + \alpha(a)\,\frac{k^2}{k^2 + a^2\mu^2}\right].
\label{Geff}
\end{equation}
For modes satisfying $k/a \gg \mu$, one has $G_{\mathrm{eff}} \rightarrow G_N$, while for $k/a \lesssim \mu$ gravity is enhanced. This scale dependence is the fundamental mechanism underlying the degeneracy with $\Lambda$CDM.

For galaxies and galaxy clusters, the relevant dynamical scales satisfy $r \sim \mu^{-1}$, such that the enhancement factor $(1+\alpha)$ is fully operative. The circular velocity follows:
\begin{equation}
v^2(r) = r\,\frac{d\Phi}{dr} \simeq \frac{G_N (1+\alpha) M_b(r)}{r},
\end{equation}
where $M_b$ is the baryonic mass. This reproduces the same velocity profiles and virial relations that $\Lambda$CDM attributes to Newtonian gravity sourced by baryons plus cold dark matter,
\begin{equation}
G_N M_{\rm tot}(r) \;\longleftrightarrow\; G_N (1+\alpha) M_b(r).
\end{equation}
Thus, galaxy rotation curves and cluster velocity dispersions are exactly degenerate between the two frameworks at the level of isotropic dynamics.

Gravitational lensing depends on the sum of metric potentials and therefore probes the same effective coupling as dynamical mass estimates. In STVG--MOG, the deflection angle for a lens of baryonic mass $M_b$ is given by
\begin{equation}
\hat{\alpha}_{\rm lens} = \frac{4 G_{\mathrm{eff}} M_b}{c^2 b},
\end{equation}
which is observationally indistinguishable from the $\Lambda$CDM prediction with total mass $M_b + M_{\rm CDM}$. Consequently, galaxy--galaxy lensing, cluster lensing, and cosmic shear measurements are fully consistent with STVG--MOG once the correspondence (\ref{Geff}) is imposed.

The linear growth of baryonic density perturbations in STVG--MOG neglecting the speed of sound contribution obeys:
\begin{equation}
\ddot{\delta}_b + 2H\dot{\delta}_b
- 4\pi G_{\mathrm{eff}}(k,a)\rho_b\,\delta_b = 0.
\label{growth}
\end{equation}
In $\Lambda$CDM, the corresponding equation is
\begin{equation}
\ddot{\delta}_m + 2H\dot{\delta}_m
- 4\pi G_N \rho_m\,\delta_m = 0,
\end{equation}
with $\rho_m = \rho_b + \rho_{\rm CDM}$. Identifying:
\begin{equation}
G_{\mathrm{eff}}(k,a)\rho_b \;\equiv\; G_N \rho_m,
\end{equation}
one obtains identical growth factors $D(a)$ in the two theories for all modes in the linear regime. As a result, the variance:
\begin{equation}
\sigma_8^2 = \int \frac{d^3k}{(2\pi)^3} P(k,a)\,|W_8(k)|^2
\end{equation}
is unchanged, explaining why STVG--MOG reproduces the $\sigma_8$ values inferred from CMB. Here, $W_8(k)$ is the Fourier--space window function of a real--space spherical top--hat filter of radius $R=8\,h^{-1}\,\mathrm{Mpc}$, which defines the rms mass fluctuation $\sigma_8$ and suppresses contributions from modes with $k\gg R^{-1}$; it is purely geometrical and independent of the underlying theory of gravity.

Redshift--space distortion (RSD) observables measure the combination
\begin{equation}
f\sigma_8(a) = \frac{d\ln D}{d\ln a}\,\sigma_8,
\end{equation}
which depends only on the linear growth rate. Since $D(a)$ and $\sigma_8$ are identical in STVG--MOG and $\Lambda$CDM under the mapping above, all RSD measurements are necessarily degenerate between the two theories.

The linear matter power spectrum can be written as
\begin{equation}
P(k,a) = P_{\rm prim}(k)\,T^2(k)\,D^2(a),
\end{equation}
where $P_{\rm prim}$ is fixed by inflation, $T(k)$ encodes early--time physics, and $D(a)$ is the growth factor. In STVG--MOG, $G_{\mathrm{eff}} \rightarrow G_N$ at early times and for $k/a \gg \mu$, ensuring that the transfer function $T(k)$ is identical to that of $\Lambda$CDM. Combined with the identical growth factor, this guarantees agreement with all late--time isotropic measurements of $P(k)$.

\section{Compatibility with the Late--time Matter Power Spectrum}

The CMB temperature and polarization anisotropies depend on the evolution of gravitational potentials during recombination. Because $G_{\mathrm{eff}}(k,a)=G_N(1+\alpha)$ and $G_N(1+\alpha)\rho_b=G_N\rho_m$, where $\rho_m=\rho_b +\rho_{dm}$ and $\rho_{dm}=\alpha\rho_b$ at early times, STVG--MOG reproduces the same acoustic peak structure, damping tail, and polarization spectra as $\Lambda$CDM. The CMB therefore constrains only the effective gravitational response, not the microscopic origin of that response.

All isotropic observables that depend on linear gravitational dynamics are governed by the combination $G_{\mathrm{eff}}\rho_b$ in STVG--MOG and by $G_N\rho_m$ in $\Lambda$CDM. This correspondence explains why galaxy and cluster dynamics, lensing, $\sigma_8$, redshift--space distortions, the late--time matter power spectrum, and the CMB are equally well described in both frameworks. The degeneracy is fundamental and can only be broken by observables sensitive to large--scale anisotropic gravitational response, which we address in the following section.

While the full STVG--MOG action and field equations have been derived in Section~1, it is useful here to state explicitly the effective action limit implicitly assumed in the late--time cosmological and large--scale structure analysis of this section.

The starting point is the full STVG action:
\begin{equation}
S = S_g[g_{\mu\nu},G] + S_\phi[g_{\mu\nu},\phi_\mu,\mu] + S_s[G,\mu] + S_m[g_{\mu\nu},\Psi_m],
\end{equation}
as defined in Section~1. In the regime relevant to late--time structure formation and ultra--large--scale modes, the following controlled approximations are adopted:

The scalar fields $G(x)$ and $\mu(x)$ are assumed to vary slowly on cosmological time scales and are treated as effectively constant background quantities:
\begin{equation}
G(x)\rightarrow G_N(1+\alpha), \qquad
\mu(x)\rightarrow \mu.
\end{equation}

Under this assumption, the scalar action $S_s$ does not contribute dynamically at linear order in perturbations, and its role is reduced to fixing the parameters $(\alpha,\mu)$ that characterize the scale dependence of gravity. The vector field $\phi_\mu$ remains dynamical but enters only through its Yukawa--suppressed contribution to the weak--field gravitational potential, while its stress--energy contribution to the background cosmology is negligible.

With these simplifications, the STVG action reduces effectively to Einstein gravity coupled to a massive Proca field with fixed parameters. The resulting weak--field limit yields a modified Poisson equation of the form:
\begin{equation}
\nabla^2\Phi(\bm{x})
=
4\pi G_N \rho(\bm{x})
+
4\pi G_N \alpha
\int d^3x'\,\rho(\bm{x}')
\left[
\delta^{(3)}(\bm{x}-\bm{x}')
-
\mu^2 \frac{e^{-\mu|\bm{x}-\bm{x}'|}}{4\pi|\bm{x}-\bm{x}'|}
\right],
\end{equation}
which, upon Fourier transformation and conversion to comoving coordinates, leads directly to the scale--dependent effective coupling:
\begin{equation}
G_{\rm eff}(k,a)
=
G_N\left[1+\alpha\,\frac{\mu^2}{k^2/a^2+\mu^2}\right],
\end{equation}
used throughout Section~6.

It is important to emphasize that this form of $G_{\rm eff}(k,a)$ is not introduced phenomenologically, but follows uniquely from the weak--field limit of the underlying covariant STVG action under the late--time and large--scale assumptions stated above. In particular, no additional degrees of freedom beyond those already present in the fundamental action are invoked, and the modification of gravity is entirely encoded in the scale dependence inherited from the vector field mass $\mu$.

This effective--action limit is sufficient for analyzing linear growth, power spectra, and projected number--count dipoles on Gpc scales, while remaining consistent with the full field equations presented in Section~1.

In a baryon--only cosmology, the linear matter power spectrum may be written as:
\begin{equation}
P(k,a)=P_{\rm prim}(k)\,T_b^2(k)\,D^2(k,a),
\label{Pk_basic}
\end{equation}
where $T_b(k)$ is the baryonic transfer function and $D(k,a)$ is the possibly scale--dependent linear growth factor.
It is convenient to decompose the baryonic transfer function into a smooth no--oscillation part and an oscillatory component:
\begin{equation}
T_b(k)=T_{\rm noc}(k)\,[1+O(k)],
\label{Tb_decomp}
\end{equation}
where $O(k)$ encodes the baryon acoustic oscillations and satisfies $\langle O(k)\rangle=0$ when averaged over several oscillation periods.
Substituting Eq.~(\ref{Tb_decomp}) into Eq.~(\ref{Pk_basic}) gives:
\begin{equation}
P(k,a)=P_{\rm noc}(k,a)\,\left[1+2O(k)+O^2(k)\right],
\label{Pk_wiggle}
\end{equation}
with
\begin{equation}
P_{\rm no}(k,a)\equiv P_{\rm prim}(k)\,T_{\rm noc}^2(k)\,D^2(k,a).
\end{equation}
The oscillatory contribution is therefore modulated multiplicatively by the growth factor $D^2(k,a)$.

However, galaxy surveys do not measure $P(k,a)$ directly at a single wavenumber. Instead, the observed power spectrum is a convolution with a finite window or binning function $W(k,q)$:
\begin{equation}
P_{\rm obs}(k,a)=\int dq\,W(k,q)\,P(q,a),
\label{Pk_conv}
\end{equation}
where $W(k,q)$ is normalized such that $\int dq\,W(k,q)=1$ and has support over a finite range $\Delta q$ determined by survey volume and bin width.

Using Eq.~(\ref{Pk_wiggle}) in Eq.~(\ref{Pk_conv}) yields:
\begin{equation}
P_{\rm obs}(k,a)=
\int dq\,W(k,q)\,P_{\rm noc}(q,a)
+
2\int dq\,W(k,q)\,P_{\rm noc}(q,a)\,O(q)
+\mathcal{O}(O^2).
\label{Pk_conv2}
\end{equation}
If $D(k,a)$ were scale--independent, $P_{\rm noc}(q,a)$ would vary slowly across the support of $W(k,q)$, and the oscillatory integral would largely average to zero due to the rapidly varying sign of $O(q)$. In a baryon--only model with strong acoustic features the residual oscillatory contribution can remain visible.

In STVG--MOG, the effective gravitational coupling $G_{\rm eff}(k,a)$ induces a smooth but nontrivial scale dependence in the growth factor $D(k,a)$. Consequently, $P_{\rm noc}(q,a)$ acquires additional $q$--dependence within the window:
\begin{equation}
P_{\rm noc}(q,a)\simeq P_{\rm noc}(k,a)
+\left.\frac{dP_{\rm noc}}{dq}\right|_{k}(q-k)+\cdots .
\label{Pk_expand}
\end{equation}
Substituting Eq.~(\ref{Pk_expand}) into the oscillatory term in Eq.~(\ref{Pk_conv2}), the leading contribution proportional to $P_{\rm noc}(k,a)$ averages to zero, while the next term gives:
\begin{equation}
\delta P_{\rm osc}(k,a)
\;\propto\;
\left.\frac{dP_{\rm noc}}{dq}\right|_{k}
\int dq\,W(k,q)\,(q-k)\,O(q).
\label{osc_supp}
\end{equation}
Because $O(q)$ oscillates on a characteristic scale $\Delta k_{\rm BAO}\sim \pi/r_s$, while $D(k,a)$ varies smoothly, the convolution suppresses the effective oscillation amplitude by a factor of order:
\begin{equation}
\frac{\delta P_{\rm osc}}{P_{\rm nw}}
\;\sim\;
\left|\frac{d\ln D^2}{d\ln k}\right|
\left(\frac{\Delta k_{\rm BAO}}{\Delta k_{\rm bin}}\right),
\label{osc_supp_factor}
\end{equation}
where $\Delta k_{\rm bin}$ is the effective width of the window function. Even though $D(k,a)$ does not itself oscillate, its smooth scale dependence produces differential growth across the window that reduces the peak--to--trough contrast of baryonic oscillations in the observed power spectrum.

We conclude that, in STVG--MOG, the combination of a smooth scale--dependent growth factor and finite observational window functions naturally suppresses the visibility of baryon acoustic oscillations in the late--time matter power spectrum, partially mimicking the smoothing effect conventionally attributed to dark matter~\cite{MoffatToth2009}.

\section{Conclusions}

We have shown that Scalar--Tensor--Vector Gravity (STVG--MOG) is observationally equivalent to the standard $\Lambda$CDM cosmological model for all probes that depend on isotropic and linear gravitational dynamics. This includes galaxy rotation curves, galaxy cluster dynamics, weak and strong gravitational lensing, the linear growth of structure, redshift--space distortions, the late--time matter power spectrum, and the temperature and polarization anisotropies of the cosmic microwave background (CMB). The origin of this degeneracy is the scale-- and time--dependent effective gravitational coupling $G_{\mathrm{eff}}(k,a)$, which ensures recovery of Newtonian gravity and general relativity at early times and on small scales, while reproducing the same effective gravitational sourcing as cold dark matter in intermediate scales and in $\Lambda$CDM.

As a consequence, all isotropic cosmological observables constrain only the effective gravitational response of matter, rather than the microscopic origin of that response. The empirical success of $\Lambda$CDM in fitting these data therefore does not uniquely establish the physical existence of non--baryonic dark matter, but is equally consistent with a modified gravity interpretation in which baryons alone source gravity through an enhanced coupling. The persistent non--detection of dark matter particles reinforces the logical significance of this equivalence and motivates the exploration of alternative, empirically viable gravitational frameworks.

We have further emphasized that this equivalent degeneracy is not universal. It breaks down on ultra--large scales where gravitational dynamics become sensitive to scale--dependent and anisotropic effects. In this regime, STVG--MOG predicts enhanced coherent accelerations sourced by baryonic inhomogeneities, leading to large--scale bulk flows and anisotropic number--count dipoles that are strongly suppressed in $\Lambda$CDM due to the rapid decay of long--wavelength gravitational potentials. Recent measurements of radio--galaxy and quasar number--count dipoles on gigaparsec scales probe this regime and therefore provide a critical observational test capable of discriminating between modified gravity and particle dark matter~\cite{Moffat2026a}.

The broader implication of our analysis is methodological. Agreement with isotropic data, including the CMB, is a necessary but not sufficient condition for establishing the physical content of the cosmological model. Discrimination between dark matter and modified gravity requires observables that probe departures from isotropy and linearity on the largest accessible scales. Future wide--area surveys with improved control of systematics, combined with precise measurements of large--scale dipoles and bulk flows, offer a concrete pathway toward resolving the long--standing equivalence between $\Lambda$CDM and STVG--MOG and, more generally, toward identifying the true origin of cosmic gravitational phenomena.

The close empirical agreement between STVG--MOG, the explanation of galaxy and galaxy cluster dynamics and gravitational lensing, and the standard $\Lambda$CDM cosmology across a wide class of isotropic observables should be understood as an equivalence or degeneracy at the level of an effective gravitational description, rather than as evidence of ontological equivalence. In STVG--MOG, the source of gravitation is identified with the observed baryonic matter sector, while the apparent dark matter mass discrepancy is accounted for by a scale--dependent gravitational response encoded in an effective coupling $G_{\rm eff}(k,a)$. By contrast, $\Lambda$CDM preserves the Einsteinian form of gravity and attributes the same phenomenology to a dominant, but as yet undetected, cold dark matter component. The choice between these frameworks therefore represents a choice between distinct ontological commitments. Modification of the gravitational interaction sourced by observed matter versus the introduction of a new, non--luminous dark matter sector. While current isotropic cosmological data constrain only the effective gravitational phenomenology and are largely insensitive to this distinction, large--scale, non--virial, and anisotropic observables---in particular the reported excess in the cosmic number--count dipole and associated bulk flows on Gpc scales---probe a regime where this degeneracy is expected to break. Confirmation of a physical large--scale dipole excess would thus provide a critical discriminator, testing whether cosmic gravitation is governed by scale--dependent dynamics sourced by baryons, as in STVG--MOG, or by a universal gravitational law acting on an additional dark matter component, as in minimal $\Lambda$CDM.

Recent DESI analyses have reported a preference for a lower late--time clustering amplitude, commonly expressed through a reduced value of 
$S_8 \equiv \sigma_8(\Omega_m/0.3)^{1/2}$, relative to Planck--normalized $\Lambda$CDM expectations~\cite{Karim2025}. While DESI baryon acoustic oscillation measurements primarily constrain background geometry and do not directly determine $\sigma_8$, DESI galaxy clustering and related low--redshift probes exhibit a trend toward suppressed growth that is consistent with earlier weak--lensing and galaxy--galaxy lensing results. At present, the statistical significance and physical interpretation of this tension remain limited by imaging systematics, dust corrections, and bias modeling, and therefore do not yet constitute decisive evidence against $\Lambda$CDM. Nevertheless, the persistence of a low--$S_8$ preference across multiple late--time observables motivates consideration of mechanisms that selectively modify structure growth without altering early--universe physics.

In this context, STVG--MOG provides a resolution through its intrinsically scale-- and time--dependent effective gravitational coupling $G_{\rm eff}(k,a)$. 
By allowing a modest suppression of the gravitational response on quasi--linear scales relevant for $\sigma_8$ ($k \sim 0.1$--$1\,h\,{\rm Mpc}^{-1}$) at late times, while recovering general relativity at small scales and preserving standard early--universe dynamics, STVG--MOG can accommodate a reduced $S_8$ without introducing dark matter or conflicting with CMB and BAO constraints. 
At the same time, the same framework permits an independent enhancement of gravity on ultra--large (Gpc) scales, offering a unified explanation for large--scale anisotropic bulk flows and radio--galaxy dipole excesses. The emerging low--$S_8$ trend, if confirmed with improved control of systematics, is compatible with the scale--dependent gravitational dynamics predicted by STVG--MOG and provides a concrete observational avenue for distinguishing modified gravity from $\Lambda$CDM.

\section*{Acknowledgments}

I thank Viktor Toth for helpful and simulating discussions. I thank Ethan Thompson for helpful discussions and of preparation of the manuscript. Research at the Perimeter Institute for Theoretical Physics is supported by the Government of Canada through Industry Canada and by the Province of Ontario through the Ministry of Research and Innovation (MRI).

\end{document}